\documentclass[iop]{emulateapj}

\slugcomment{Not to appear in Nonlearned J., 45.}

\shorttitle{The dominant epoch of star formation in the Milky Way}
\shortauthors{Snaith et al.}

\begin{document}


\title{The dominant epoch of star formation in the Milky Way formed the Thick Disc}


\author{Owain Snaith\altaffilmark{1,2}, Misha Haywood\altaffilmark{1}, P. Di Matteo\altaffilmark{1}, 
Matthew D. Lehnert\altaffilmark{3}, Fran\c coise Combes\altaffilmark{4}, David Katz\altaffilmark{1} and Ana G\'omez\altaffilmark{1} }

\email{owain.snaith@obspm.fr}


\altaffiltext{1}{GEPI, Observatoire de Paris, GEPI, CNRS, Universit\'e Paris Diderot, 5 Place Jules Janssen, 92190 Meudon, France}
\altaffiltext{2}{present address: Department of Physics \& Astronomy, University of Alabama, Tuscaloosa, Alabama, USA}
\altaffiltext{3}{Institut d'Astrophysique de Paris, UMR 7095, CNRS, Universit\'e Pierre et Marie Curie, 98 bis Bd Arago, Paris, France}
\altaffiltext{4}{Observatoire de Paris, LERMA, CNRS, 61 Av de l'Observatoire, 75014, Paris, France}

\begin{abstract} 
We report the  first robust measurement  of the
Milky Way  star formation history  using the imprint left  on chemical
abundances of  long-lived stars. The  formation of the  Galactic thick
disc occurs  during an  intense star formation  phase between  9.0 (z$\sim$1.5) and
12.5 Gyr (z$\sim$4.5) ago and is followed by a dip (at z$\sim$1.1) lasting about 1 Gyr. Our results
imply that the thick disc is  as massive as the Milky Way's thin disc,
suggesting a fundamental role of  this component in the genesis of our
Galaxy, something that had  been largely unrecognized. 
This new picture implies that huge quantities of gas  necessary
to feed the building of the thick disc must have been present 
at these epochs, in contradiction with the long-term infall  
assumed by chemical evolution models in the last two decades.
These results allow us
to fit the Milky Way within  the emerging features of the evolution of
disc galaxies in the early Universe.
\end{abstract}





\addtocounter{footnote}{-5}
\section{Introduction}

Galaxies  in   the  early  Universe  have   clumpy,  highly  turbulent,
gas-dominated  discs \citep{elm07, gen08} and  are  vigorously forming  stars. They  are
believed  to  sustain  their  high  rates of  star  formation  by  the
accretion   of   cold  gas   along   filaments   into  their   central
regions \citep{ker05, ker09, dek09}. The most intense phase  of star formation takes place at
redshifts  ~1-5 \citep{hop06}, with  steep decline  thereafter \citep{mad96}. It  is currently
unknown  how our Milky  Way (hereafter MW) galaxy  fits into  this general  scheme. A
fundamental piece of the puzzle is missing in our understanding of the
star  formation history  (hereafter SFH) of the  MW,  and its  association with
galaxies observed in the early Universe.  The aim of this study
is to present the first results of a new method to measure the SFH in our Galaxy using chemical abundances.

The chemical elements formed
and ejected by the early  generations of stars during the formation of
the  MW are  now found  in the  atmospheres of  old, long-lived
stars. The  absolute and relative  abundances of these elements  are a
gauge  of the  past star  formation  as the  Galaxy evolved.  Observed
variations in  the elemental abundance  patterns as a function  of the
age of the stars can be  reproduced with a chemical evolution model to
infer the SFH of our Galaxy. 
The method has the huge advantage that, if stars have different origins in the disc,  
and follow a uniform
evolution (as is the case for the thick disc, see below),
the measured SFH is non-local, and valid for the entire population.
Although the approach
is  model-dependent, and its validity relies on assumptions which are still largely debated - such as the 
invariance of the IMF with time - it  is much  more economical  and straightforward
than  standard methods,  which aim  to reconstruct  the  Galactic SFH  by determining  individual stellar ages  for volume
complete samples. 
Our method does not rely on star density, it uses only abundance variations 
as a function of age, thus is not limited by number densities-related 
problems such as the design of volume complete samples, scale height 
corrections, etc. 
Moreover, it gives access  to detailed variations of the SFH at ancient epochs, a domain that is likely to remain
inaccessible with  standard methods at  least until the advent  of the ESA
astrometric mission, Gaia.  Counting  stars as a function of age for ancient populations 
is indeed impractical for the time being, simply because there are not enough old stars with
sufficiently  accurate ages.

\section{Observations} 

We base our study on a spectroscopic dataset of 1111 F, G and K dwarfs
stars from the solar vicinity \citep{adi12}. Ages were obtained using
a Bayesian estimate of  the probability distribution function for each
star \citep{hay13} and a  severe pruning  to only include stars with robust
age  estimates has been applied. Moreover, in the present study we exclude outer thin disc stars 
(stars on the thin disc sequence -- e.g, low-alpha stars, see \cite{hay13} -- 
with  [Fe/H]$\le$ -0.2 dex), because of their possible separate chemical evolution history \citep[][]{hay13}. This reduces  the  sample to  281  stars. Fig.  1 shows  the distribution  of   [Si/Fe]   versus  age for the final sample.  It illustrates a change of slope  in the declining ratio of [$\alpha$/Fe] at  about 8  Gyr.  Two  distinct phases  in  the abundance  enrichment
history  of the  MW are clearly  evident \citep{hay13}.

As in \citet{hay13}, we {\it define} the thin and thick discs as corresponding to these two different phases, 
with a transition between these two formation epochs about 8--9 Gyr ago.  
Starting  from  this   result,  we  sought  to  model  the
corresponding  variation  in  the  SFH  needed  to reproduce these trends in [$\alpha$/Fe] with time.

\begin{figure}
\includegraphics[trim=100 140 40 280,clip,width=10.cm]{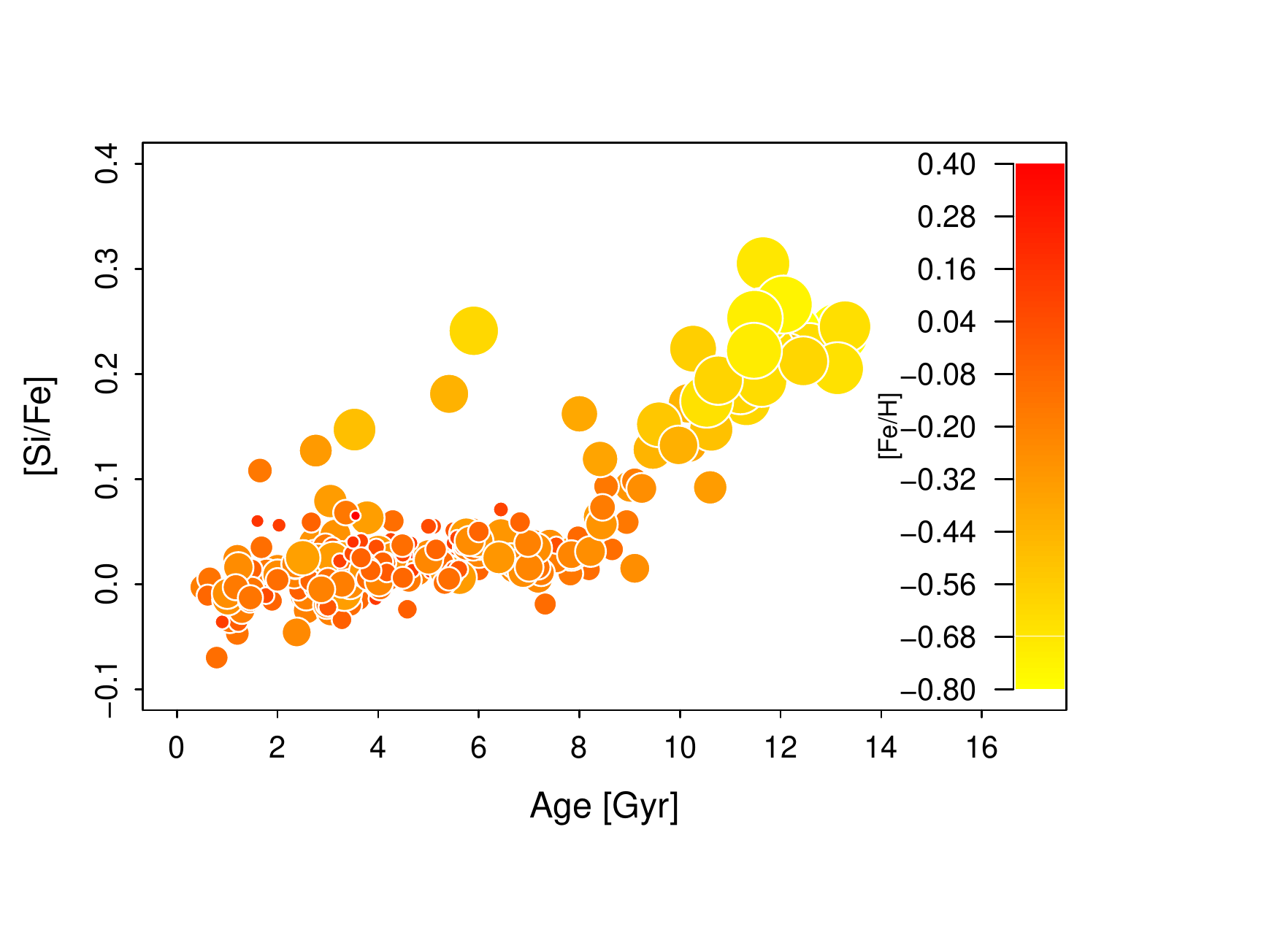}
\caption{The age-[Si/Fe] distribution of stars at the solar vicinity. Both the size and the colour code the metallicity of stars, with the colour coded according to the vertical scale. The selected stars show two regimes of [Si/Fe] evolution as a function of time, defining the thick (ages greater than 8 Gyr) and thin (ages smaller than 8 Gyr) disc populations \citep{hay13}.}
\label{alphafehage}
\end{figure}

\section{Model}

Our modelling  approach assumes the following:\\
\emph{(i)}  All  the gas is in  the system at
the  beginning,  with no inflow or outflow.  This is  similar  to  a  closed-box model \citep{tin80},  but  is
equivalent  to  saying that  most  accretion  in  the inner disc (R$<$10 kpc)
occured  early.\\
\emph{(ii)} The ISM  is
initially  devoid  of  metals   and  remains  well  mixed  during  its whole
evolution. \\
\emph{(iii)} Because  our  aim  is to  recover  the Galactic SFH, no coupling  between the star formation rate  and the gas content
is  assumed. Instead, we determine the SFH by fitting our model to the data.

Our approach  is otherwise  typical  of chemical  evolution models.
We adopt standard  theoretical yields \citep{iwa99, nomoto06, karakas10}, an IMF \citep{kro01} independent of time, and  a stellar
mass-lifetime relation dependent on the metallicity \citep{rai96}. No instantaneous recycling approximation has been implemented.

We  search  for  the  SFH that best fits the [$\alpha$/Fe]-Age relation (in this case
the alpha element used is  the Silicon), rather than [$\alpha$/Fe]-[Fe/H], as
is usually done. Our main reason for doing this is that different star
formation histories can produce similar chemical [$\alpha$/Fe]-[Fe/H] tracks,
thus  the   solution  is  degenerate   in  this  plane,   without  age
information.  Moreover, the  evolutionary [$\alpha$/Fe]-[Fe/H]  track  of the
thin  disc  being very  limited \citep{hay13},  it is  not  possible  to use  this
sequence to constrain its  detailed evolution.

In practice, at t = 0, the system contains  a pristine gas of mass 1. For each time
step  the  code looks  at  the  provided SFH and
calculates  the amount of  mass that  is converted  from the  ISM into
stars and  subtracts it from the  ISM. The mass  of O, Mg, Fe,  Si, H,
corresponding to the abundance of the ISM at that time is also removed
from  the ISM and  used to  calculate the  metallicity of  the stellar
population.

The code then  looks at the stars created in  every previous time step
and calculates  the amount of gas  and metals that is  returned to the
ISM. This is done by  interpolating the cumulative yields produced for
all  stars at  the end  of  their lives  and taking  into account  the
metallicity of the  star at its formation. This mass  is then added to
the ISM and the fraction of  recycled gas is removed from the star and
returned to the gas component. A chemical track is finally generated resulting
from this particular SFH, and fitted to the data through a $\chi^2$ approach, 
using the difference between the model value of [Si/Fe] at the age
of each star in the data and the observed [Si/Fe]. 
The data is fitted iteratively. 
At each iteration an entire SFH is created and the chemical evolution
code is used to calculate the corresponding chemical track. 
The algorithm requires an initial guess SFH and a user supplied convergence criterion. 
A series of best fits are found by bootstraping the sample, and these best fits 
are averaged in order to smooth the SFH (see Snaith et al. (2014)
for a detailed description).

\begin{figure*}
\includegraphics[trim=100 140 80 280,clip,width=9.5cm]{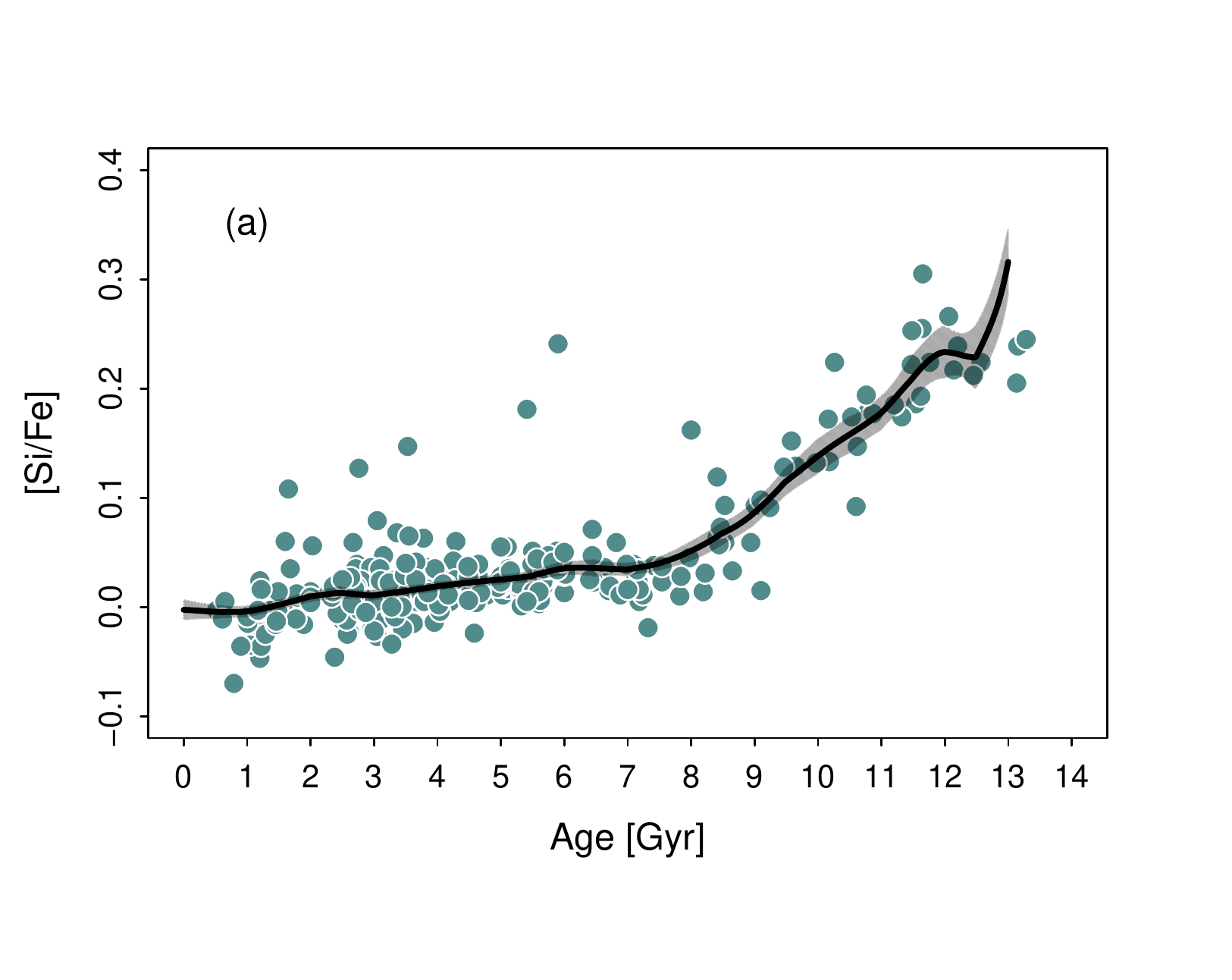}
\includegraphics[trim=100 140 80 280,clip,width=9.5cm]{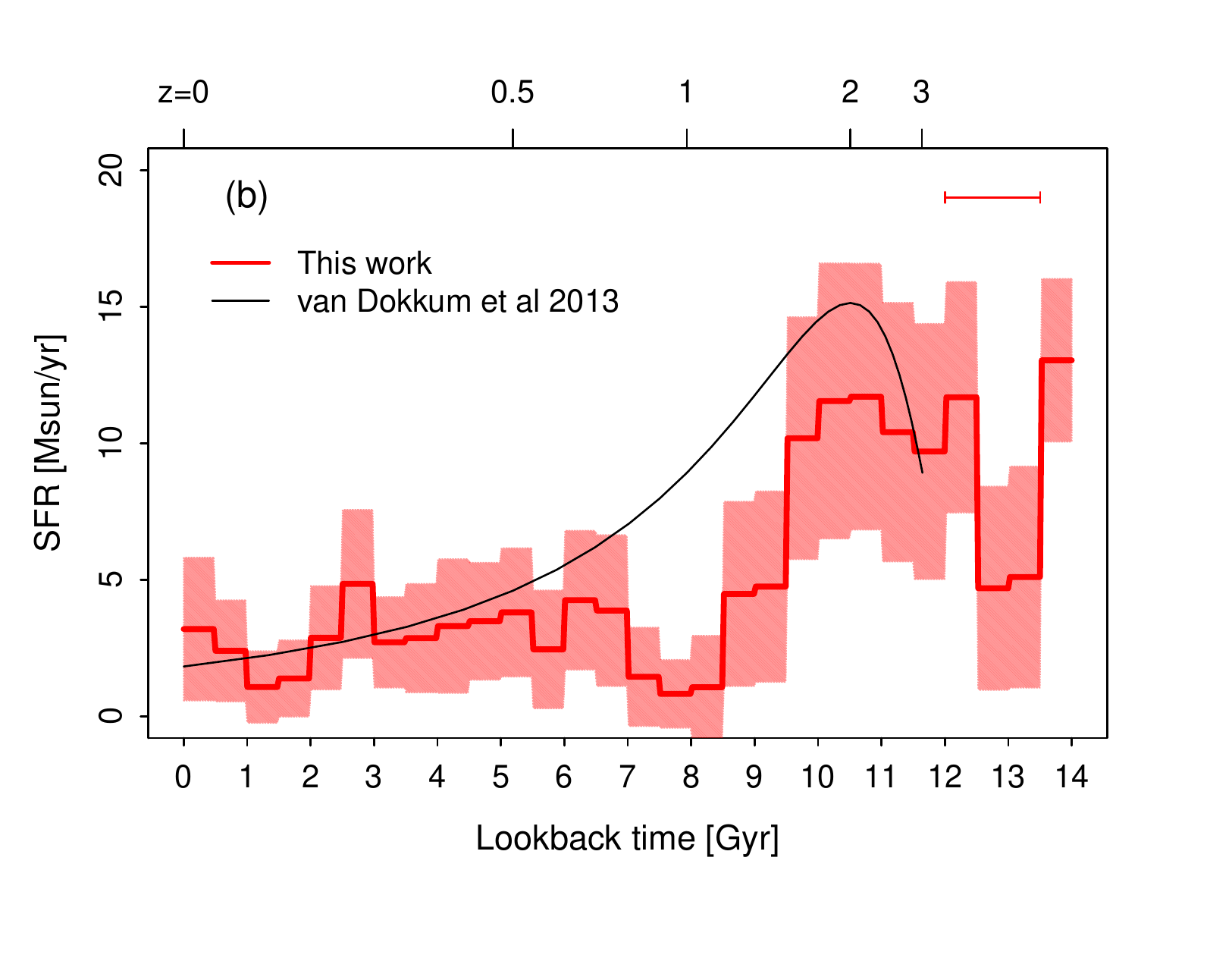}
\includegraphics[trim=100 140 80 300,clip,width=9.5cm]{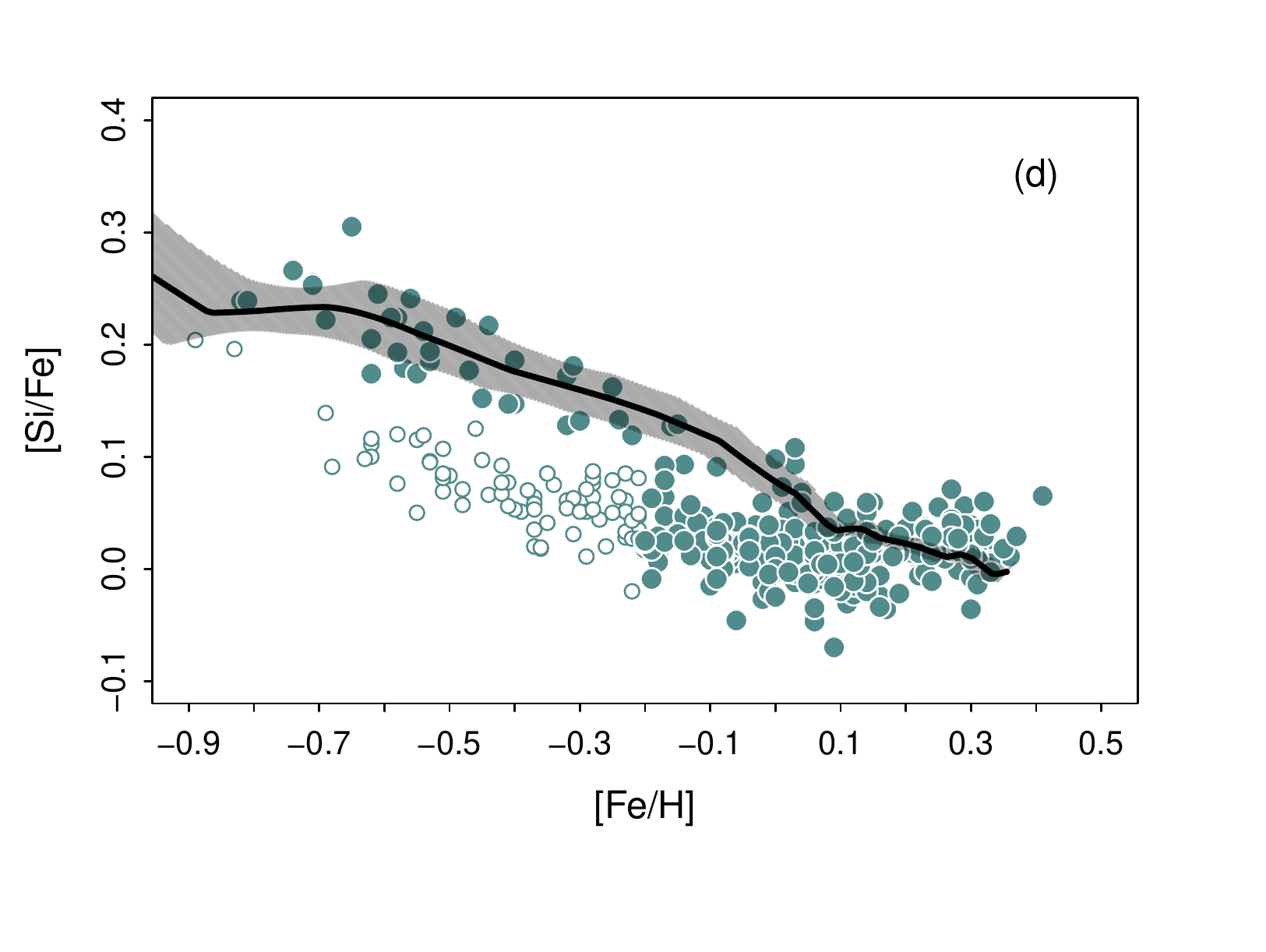}
\includegraphics[trim=100 140 80 300,clip,width=9.5cm]{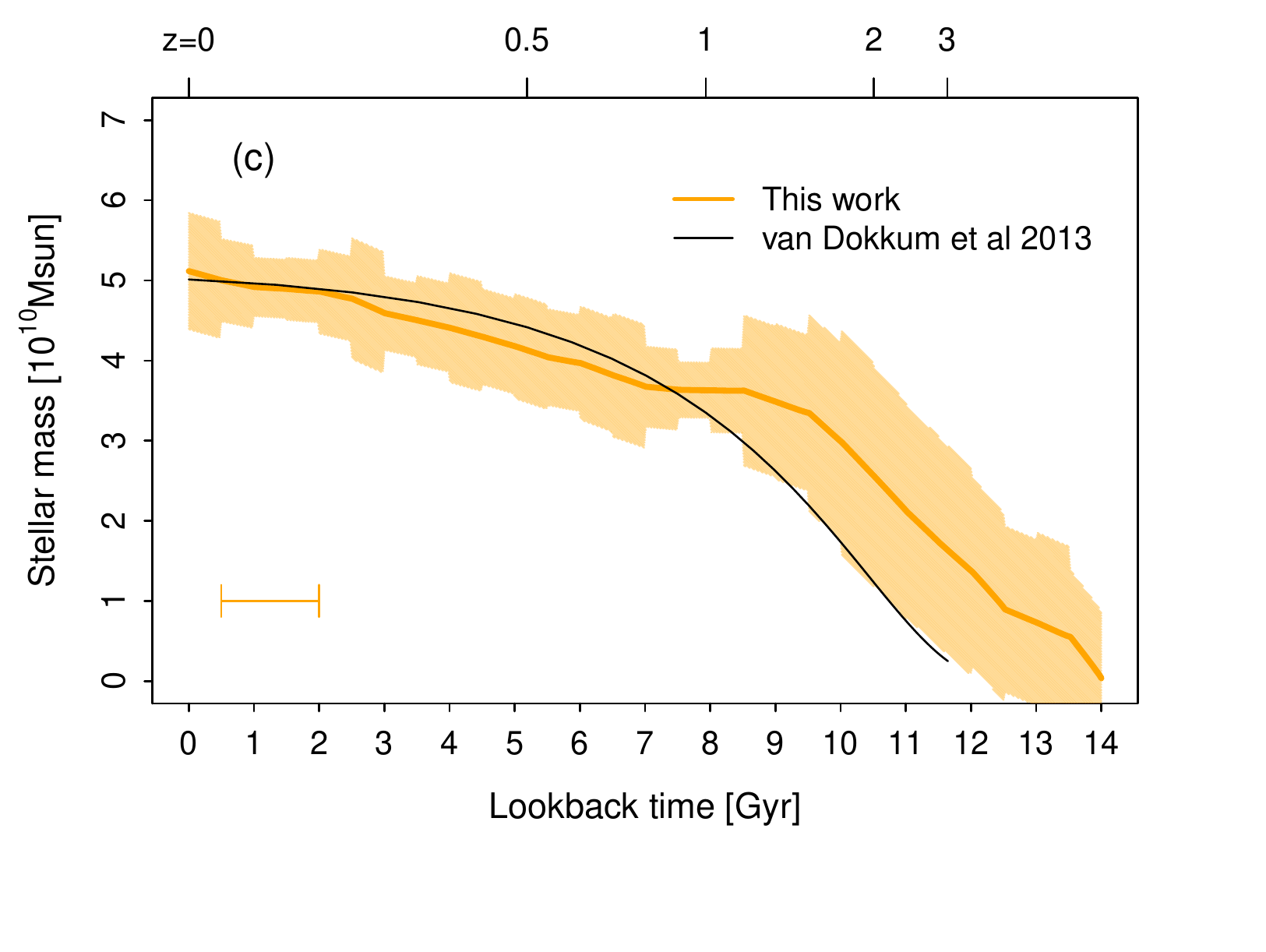}
\caption{\emph{Panel~(a): } [Si/Fe] versus age relation (black curve) as determined by fitting our model to (inner disc) data from Adibekyan et al (2012) and Haywood et al. (2013; green circles). The grey area is the 1-sigma error estimated by a bootstrapping technique. \emph{Panel~(b): } Star formation rate variation as a function of age. The SFR is normalized so that the total mass of stars formed in the disc amounts to 5.10$^{10}M\odot$. The red region gives 1-sigma error estimated through bootstrap technique. The horizontal error bars in the figures show the estimated errors on ages \citep{hay13}. The black curve is the SFH as a function of redshift estimated by selecting Milky
Way-like galaxies via halo abundance matching (van Dokkum et al. 2013).  The SFR is found by minimising the distance between the model chemical track and the variation of abundances with age given in Panel~(a). \emph{Panel~(c): } Cumulative stellar mass as a function of redshift for the MW from our analysis (yellow curve and shaded region) compared to that of Milky Way-analog galaxies from van Dokkum et al. (2013; black curve). We estimated the cumulative stellar mass by integrating the SFH shown in Panel (b). \emph{Panel~(d): } [Si/Fe]-[Fe/H] chemical evolution track obtained by averaging a hundred best fits to the data in the [Si/Fe]-Age plane with 1-sigma error distribution (dark shaded regions) estimated by a bootstrapping technique.}
\label{results}
\end{figure*}

\section{Results}  \label{results}

The result of the fitting procedure to the Age-Silicon data is given in Fig.~2(a), with the corresponding 
SFH in Fig.~2(b). The star formation obtained suggests that:\\
 (1) the Galaxy underwent an intense phase
of star  formation between 9 and 13 Gyr ago, during  the formation of  the thick disc. At those epochs, the SFR peaked at 3--4 times the value characterising the subsequent thin disc formation phase;   \\ 
(2) this intense SF phase lasted $\sim$ 4 Gyr and produced about as much mass in stars as that produced in the following 8 Gyr, during the
thin disc formation (Fig.~2(c)). In other words, the thick disc of the Galaxy is about as
massive  as the  thin disc  (in agreement  with recent  mass estimates
using  structural  parameters, see Sect.~5.1); \\
(3)  the estimated  SFH shows  a distinct  dip  at $\sim$ 8-9 Gyr. A statistical analysis (Snaith et al., 
2014) indicates that this dip is  robust, and implies that  the end of the  thick disc formation
phase is marked by a decrease in the star formation.

A caveat must be added. As mentioned in the introduction, although our sample is local ($<$100pc), as long as 
the analysis is based on stars with sufficiently large and different eccentricities, the recovered SFH 
is non-local, and is valid for the whole population. While this is true for the thick disc, it becomes less 
correct for younger stars that have orbits more confined to the solar circle. 
In this respect, it is difficult to ascertain if 
the dip in the SFR at 7-9 Gyr is a local feature or not. 
Reciprocally, stars that avoid the solar circle will not be represented in the Age-abundance plane, 
and not taken into account in the reconstructed SFH. 
From that point of view, our position, at the limit between the inner and outer discs, has an advantage:
the solar vicinity is a fair sampling of all known populations of the disc, while we know that the inner disc lacks the metal-poor
thin disc stars, and the outer disc lacks both the metal-rich alpha-poor thin disc and the alpha-rich metal-poor
thick disc.


Finally, 
it is interesting to show the chemical evolution in the [Si/Fe]-[Fe/H] plane generated by the recovered SFH. This is done in Fig.~2(d), which illustrates the [Si/Fe]-[Fe/H] chemical track of the model 
that results from averaging individual [Si/Fe]-[Fe/H] tracks. Each of these tracks has been derived using the best fit SFH solution obtained by fitting the Age-[Si/Fe] relation through bootstraping the data. It is striking to see that, even if the model has not been designed to fit this relation, the obtained chemical tracks describe very well the thick disc sequence. 
In contrast, the thin disc is less well described, as expected, since the mean [$\alpha$/Fe] and metallicity vary very
little during the thin disc phase, while the dispersion in metallicity  is very large at any given age (or [$\alpha$/Fe]), due to the
position of the Sun at the transition between the inner metal-rich and the outer metal-poor discs.

Overall, the recovered SFH allows us to match, at the same time, the Age-[Si/Fe] and the [Si/Fe]-[Fe/H] relations, 
thus guaranteeing that it is a robust description of the chemical evolution of stars observed today at the solar vicinity.

\section{Discussion} 

\subsection{The thick disc mass}
These results lead to a fundamental revision of the
role played by  the thick disc in galactic  evolution, 
as already suggested by \cite{ber01}, and more recently, \cite{fuh11, fuh12}.
With the recent recognition that our MW's stellar mass is dominated by its disc
- the contribution  of the central spheroidal component  being 10\% at
most \citep{she10, kun12, dim13} -  the thick  disc is arguably  the dominant  old stellar
population of  the MW. 

The most  recent thick disc  scale length measurements \citep[][but see Jayaraman et al. (2013)]{bov12a, ben11}  show that
this  population  has a  much  shorter  scale  length than  previously
thought, of about  2.0 kpc, while the thin disc scale  length is $\sim$3.6
kpc. From the stellar surface  densities derived from the SEGUE survey
data \citep{bov12b},  and  assuming that  the  thick  and  thin discs  separate  at
[$\alpha$/Fe]=0.25   dex \citep[on the SEGUE scale, ][]{bov12a},    the   thin   disc   contributes    to  $\sim$21
M$_{\odot}$.pc$^{-2}$ to  the local  stellar surface density,  and the
thick  disc  to  $\sim$8  M$_{\odot}$.pc$^{-2}$.  Correcting  for the  scale
length effect assuming the values  just quoted, we find that the thick
disc represents  $\sim$47\% of the  stellar mass within 10~kpc of the galactic
centre, in  accordance with estimates from the  SFH
given here. 

In other words,  the stellar  population responsible  for the  increase  of the
metallicity in  the Galaxy from halo  to thin disc  values is massive 
\citep[see also][for a similar conclusion]{fuh12},
and essentially concentrated in the inner Galaxy.
This is confirmed by the first APOGEE results \citep[][Fig. 14]{anders13}, where it
is shown that the MDF for stars with median galactocentric
radii between 4 and 7~kpc has a mode at [Fe/H]$\sim$ -0.2 dex, with roughly equal amount of stars above and below 
this value. 
 Interestingly, the distribution of alpha abundance as a function of metallicity given in this same plot shows that below  [Fe/H]$\sim$-0.2 dex, most stars are alpha-enhanced,
meaning that they are thick disc objects.

\subsection{The  G-dwarf problem and the accretion history of the Galaxy}

As well as a major contributor to the total mass budget of the Galaxy,  the thick disc also  
produced as much  metals as the thin disc in the following 8 Gyr of galactic evolution.

This role has been largely over-looked in chemical evolution models of
our Galaxy \citep[all models based on long term infall, e.g.][]{naa06, sch09, mic13}, which were built  on the assumption of an apparent
lack  of intermediate  metallicity  disc  stars in  the  Galaxy -  the
so-called G-dwarf problem \citep{van62} - simply because these stars are poorly
represented in the solar neighbourhood. A consequence is that Galactic
chemical evolution models so far have underestimated the importance of
this population, leading  to a biased view of  Galactic evolution. The
G-dwarf   problem as such vanishes because (1) the local metallicity distribution function (MDF) is strictly local and 
cannot be used to constrain models that aim to give a general description of the chemical evolution of the 
discs\footnote{This is applicable when comparing the (relative) frequency of stars of local samples to models
that are designed to describe the chemical evolution of the entire disc. Our sample is local as well,  
but we use variations of metallicities and abundances as a function of age in our derivation of the SFH, not their number densities.} and (2) stars of intermediate metallicities are  in  fact  largely present  in  the Galaxy  and
correspond to the thick disc.

This is of crucial  importance also for the long-term evolution
of the Galaxy, since it  alleviates the need for prolonged infall onto
the inner  ($R <$10~kpc) disc over the  last 8 Gyr. Over  this period, the
MW may  have sustained  its star  formation from  the  gas not
consumed during  the thick disc  formation epoch and by  recycling the
mass lost from disc stars \citep{lei11}.

 Such  a  picture   is  not  in  accordance  with   models  of  galaxy
formation. We find  here that there may have been  a period of intense
accretion very early in the history of the Galaxy but, after that, the
MW may have evolved mostly by reprocessing the material lost from stars,
and by consuming  the gas left at the end of  the thick disc formation
phase.  There is  very little  need for  significant  cosmological gas
accretion beyond  the early phases  of the MW  formation.  This
may call  for a substantial  revision of our galaxy  evolution models,
which suggest that 
slow,  long-term accretion of nearly primordial gas is continuously  needed 
in the disc \citep{naa06, sch09, mic13, frat13}.

Note that these results are derived from inner disc stars, 
and are not applicable to the outer disc, which could have followed a different chemical evolution \citep{hay13}
resulting from a mixture of gas ejected by outflows from the forming thick disc, and pristine accreted gas.
In this picture,  it is expected that the first (z$>$2) gas accretion phase mainly fuels the building of the inner regions of the disc,
while significant accretion may continue in the outer regions afterwards.


\subsection{The extragalactic context}

The evolution of  our Galaxy derived from these  results fits
well with  observational studies  of distant galaxies,  which indicate
that MW  progenitors assembled more than half  of their stellar
mass  within  5  Gyr  after  the  Big  Bang \citep{lei12,van13, muz13, pat13}.  
Figure~\ref{results} (panels (\emph{b}) and (\emph{c})) illustrates the evolution of the SFR and cumulative stellar mass as a function of age compared to the one obtain by \citet{van13} for
Milky Way-type progenitors at different redshifts.  Given possible systematic effects
in the age scale, good qualitative agreement is obtained between these estimates and our recovered SFH from solar vicinity data.
These  direct
observations,  together with  the properties  of local  disc galaxies,
which  do  not  show   any  presence  of  substantial  old,  classical
spheroids \citep{kor10, fis10}, suggest  that the stellar  mass formed at  these early
epochs was likely in thick discs. Moreover, spectroscopic observations
of  distant  galaxies  show   that  their  discs  have  high  velocity
dispersions  in their  gas, similar  to that  measured in  the present
stellar thick  disc of the  MW \citep{swi11, leh13}. For these  reasons, the
formation  of the  thick stellar  disc is  a fundamental  part  of the
formation of the MW, and disc galaxies generally.

\section{Conclusions}

We have presented the  first robust measurement  of the
MW  star formation history  using the imprint left  on chemical
abundances of  long-lived stars at the solar vicinity.
Our results show that the  thick disc  formed  coincident with  the  maximum star  formation
activity  in  the Universe, between 12.5 and 9 Gyr ago, and  play  a  fundamental role  in  the
chemical evolution  of the Galaxy, as well as in its global mass budget.

The implication of these results are far reaching. Firstly, although we have no direct clue of what would be a MDF representative of
the entire disc, the fraction of stars of intermediate metallicity is significantly
higher than described by the local MDF. 
A simple closed-box MDF with a maximum at [Fe/H]$\sim$-0.1 dex
has a median at [Fe/H]$\sim$-0.3 dex, which is about the metallicity
transition between the thick and thin discs (see the age-metallicity relation in \cite{hay13}), 
and this is in accord with the equal mass balance measured here between the two discs. 
It suggests that the simple closed-box may be a good first order model to describe the  
chemical evolution of the discs. 
 Secondly, the difference in scale length between the two discs, 
if confirmed, implies that the solar vicinity MDF is strictly local and {\it cannot} 
be used as it is to infer a general chemical evolution of the Galactic disc. 
Thirdly, these results suggest that large amounts of gas must have been
available early in the Galaxy to sustain the formation of the thick disc, adding support to the choice 
of adopting a closed-box model to approximate its chemical evolution. We find   that there may have been  a period of intense
accretion very early in the history of the Galaxy, and that, by z$\sim$1, the
MW may have evolved mostly by reprocessing the material lost from stars,
and by consuming  the gas left at the end of  the thick disc formation
phase. This finding is at odds with current galaxy  evolution models,
which suggest that slow,  long-term accretion of nearly primordial gas is continuously  needed  
in the MW disc to sustain its star formation.

\acknowledgments
We wish thank the referee, Klaus Fuhrmann, for his prompt report and constructive comments.
The authors acknowledge the support of the French Agence Nationale de la Recherche (ANR) under contract ANR-10-BLAN-0508 (GalHis project).

\clearpage



\end{document}